\documentstyle[psfig,prb,aps]{revtex}
\begin{document}

\headheight-1.0cm
\headsep0.0cm
\topmargin0.6cm
\textheight24cm

\draft

\twocolumn[\hsize\textwidth\columnwidth\hsize\csname
@twocolumnfalse\endcsname

\title{Towards a first-principles theory of surface thermodynamics and
kinetics}
\author{C. Stampfl,$^{1}$ H. J. Kreuzer,$^{2}$ S. H. Payne,$^{2}$
H. Pfn\"{u}r,$^{3}$ and M. Scheffler,$^{1}$}
\address{$^{1}$Fritz-Haber-Institut der Max-Planck-Gesellschaft,
Faradayweg 4-6, D-14195 Berlin, Germany\\
$^{2}$Department of Physics, Dalhousie University, Halifax,
Nova Scotia, B3H 3J5, Canada\\
$^{3}$Institut f\"{u}r Festk\"{o}rperphysik, Universit\"{a}t Hannover,
Appelstrasse 2, D-30176 Hannover, Germany}
%\date{\today}
\maketitle

\begin{abstract}
Understanding of the complex behavior of particles at surfaces
requires detailed knowledge of both macroscopic and microscopic 
processes that take place; also certain processes depend critically 
on temperature and gas pressure. To link these processes 
we combine state-of-the-art microscopic, and macroscopic
phenomenological, theories. We apply our theory to the O/Ru\,(0001) system and
calculate thermal desorption spectra, heat of
adsorption, and the surface phase diagram. The agreement with experiment
provides validity for our approach which thus identifies the way for a
predictive simulation of surface thermodynamics and kinetics.
\end{abstract}

\vskip2pc]

The study of the physical and chemical processes that take place at
gas-surface interfaces have long been an area of intense research. This
interest is both fundamental as well as driven by the possible discovery of
important technological applications, e.g. in the field of heterogeneous
catalysis, corrosion, etc.~\cite{review,review2}. With respect
to the field of the {\em theory} of adsorption of gases on solid surfaces,
advancement in recent years has developed in two distinct, albeit
complimentary directions: (i) electronic structure calculations, at best
done by density-functional theory (DFT), to determine the geometries,
energetics, and vibrational properties of adsorbate covered surfaces, and
(ii) phenomenological models, both for the thermodynamics and the
kinetics~\cite{kinrev} of the adsorbate. If one can assume that the geometry
of the
solid surface does not change dramatically and that adsorption occurs at
well defined sites, one frequently employs a lattice gas model. A number of
parameters enter such a model such as the binding energies and vibrational
frequencies of a single adparticle in the various adsorption sites, and
their mutual lateral interactions with adparticles in close-by sites.
Traditionally, these parameters are adjusted in the theory to fit a variety
of experimental data such as phase diagrams, heats of adsorption, infrared
spectra and thermal desorption data, etc. Such an approach, while useful, is
clearly not necessarily predictive in nature, nor the parameters unique, and
may not capture the physics of the microscopic processes that are behind the
``best-fit'' adjusted ``effective'' parameters.

In the present letter, with the aim to improve upon this approach, we
combine state-of-the-art procedures of (i) microscopic theories, i.e. DFT
electronic structure calculations and (ii) macroscopic phenomenological
approaches, i.e. lattice gas and rate equations, and
Monte Carlo schemes. On doing this, we
present a consistent first-principles based approach for calculation of the
thermodynamic and kinetic properties of an adsorbate, such as heats of
adsorption, temperature programmed desorption (TPD) spectra, and the surface
phase diagram. We have chosen the system of oxygen at Ru\,(0001) for which
detailed structural~\cite{lindroos,P1,stampfl2,kostov,over,menzel},
thermodynamic \cite{pdp}, and kinetic data \cite{madey,surnev} exist. We
will show that with the present approach, a realistic description of these
physical properties is indeed feasible.

We like to mention that use of results of {\em ab initio} electronic
structure calculations as input to statistical mechanical computer codes
have also been developed for the description of bulk alloy phase
diagrams~\cite{fontaine,zunger} and island growth on crystal
surfaces~\cite{scheffler}.
To our knowledge, however, there has been no application of such an
approach for the processes that we are interested in here.

The electronic structure calculations~\cite{bockstedte} are
performed using DFT and the generalized gradient
approximation (GGA) for the exchange-correlation 
functional~\cite{perdew} (hereafter denoted as DFT-GGA).
We use the pseudopotential~\cite{troullier,martin} plane wave method and the
supercell approach to model the surface.
The position of the O atoms and the top two Ru layers 
are fully relaxed.
The DFT-GGA Ru pseudopotential yields a bulk hcp-fcc energy difference
of $\approx -0.072$~eV in good agreement with all-electron
calculations.~\cite{skriver}
For further details we refer to Refs.~\cite{stampfl2,stampfl1}

To set up a lattice gas model for O on Ru\,(0001) we require a hamiltonian,
which we express as:
\begin{equation}
\begin{array}{l}
H=E_{s}^{\rm hcp}\sum_{i}n_{i}+ E_{s}^{\rm fcc}\sum_{i}n_{i}+
\frac{1}{2}(V_{\rm 1n}^{\rm hcp}\sum_{i,a}n_{i}n_{i+a} \\
+V_{\rm 1n}^{\rm fcc}\sum_{i,a}n_{i}n_{i+a}
+V_{\rm 1n}^{\rm hcp-fcc}\sum_{i,a^{\prime}}n_{i}n_{i+a^{\prime}} \\
+V_{\rm 2n}^{\rm hcp} \sum_{i,b}n_{i}n_{i+b} +V_{\rm 2n}^{\rm fcc}
\sum_{i,b}n_{i}n_{i+b} \\
+V_{\rm 2n}^{\rm hcp-fcc} \sum_{i,b^{\prime}}n_{i}n_{i+b^{\prime}}+
V_{\rm 3n}^{\rm hcp} \sum_{i,c}n_{i}n_{i+c} \\
+V_{\rm 3n}^{\rm fcc} \sum_{i,c}n_{i}n_{i+c} +V_{\rm 3n}^{\rm hcp-fcc}
\sum_{i,c^{\prime}}n_{i}n_{i+c^{\prime}} \\
+ V_{\rm trio}^{\rm hcp}\sum_{i,a,a^{\prime
\prime}}n_{i}n_{i+a}n_{i+a^{\prime \prime}}
+ V_{\rm trio}^{\rm fcc}\sum_{i,a,a^{\prime
\prime}}n_{i}n_{i+a}n_{i+a^{\prime \prime}}+
\ldots) 
\end{array}
\end{equation}
The different unit cells of the substrate surface
are labeled by an index $i$ with $i+a$, etc. labeling
neighboring cells, and we introduce occupation
numbers $n_{i}$ = 0 or 1 depending on whether a site in cell $i$ is empty or
occupied. 
Overcounting of cells is excluded in the summations.
Eq.~1 includes consideration of hcp and fcc sites. The indicies
$a^{\prime}$, $b^{\prime}$, and $c^{\prime}$ indicate that the first,
second,
and third neighbor distances between atoms in hcp and fcc sites are
different to when they occupy the same type of sites. Here
$E_{s}^{{\rm hcp}}=|V{}_{0}|-k_{B}T\ell nq{}_{3}$ is the binding energy
of an isolated
particle in an hcp site. $|V_{0}|$ is the depth of the adsorption
potential with reference to the energy
of a gas phase molecule which adsorbs dissociatively.
$q{}_{3}$ is the partition function of the atom
accounting for its
vibration perpendicular, and its frustrated translation parallel, to the
surface. $V_{\rm 1n}^{{\rm hcp}}$, $V_{\rm 2n}^{{\rm hcp}}$ and $V_{\rm 3n}^{{\rm hcp}}$
are the first, second and third neighbor interaction energies between two
adsorbed O atoms in hcp sites. The analogous terms labeled ``fcc'' represent
the same quantities but for adsorption in fcc sites. Terms labeled
$V^{{\rm hcp-fcc}}$ represent the interactions
between atoms in hcp and fcc sites. Trio (and higher
cluster) interaction energies, $V_{{\rm trio}}$, account for additional
modifications
because the interaction between two adsorbed O atoms is changed,
when a third adatom is close by.
In fact, depending on the angles and distances, there are three different
trio interactions taken into account.

We determine the lateral interaction energies required in Eq.~1 from
DFT-GGA calculations of ordered structures of O on Ru\thinspace (0001)
(see Fig.~1). The
adsorption energy of a {\em single} oxygen atom on Ru\thinspace (0001) is
obtained using a $(3\times 3)$ structure with coverage $\theta
=1/9$. With this large O-O separation (8.26~\AA ),
which  corresponds to the fifth nearest neighbor distance of alike sites,
lateral interactions are negligible.
The
adsorption energy, with respect to 1/2O$_{2}$, is then expressed as,
\begin{equation}
E_{a}^{\theta =1/9}=E_{{\rm total}}^{{\rm O/Ru}}-
E_{{\rm total}}^{{\rm Ru}}-1/2E_{{\rm total}}^{{\rm O_{2}}}\quad .
\end{equation}
Here $E_{{\rm total}}^{{\rm O/Ru}}$, $E_{{\rm total}}^{{\rm Ru}}$, and
$E_{{\rm total}}^{{\rm O_{2}}}$ are the total energies of the
$(3\times 3)$-O/Ru\thinspace (0001) system, the clean Ru surface, and a free
O$_{2}$
molecule, respectively. Expressions analogous to Eq.~2 have been used to
derive the adsorption energies for the other coverages (see Tab.~I). We
expand the adsorption energies in terms of  two- and
three-body interactions.
The interaction parameters are derived from the
equations:
\[
\begin{array}{l}
E_{a}^{\theta =2/9}=V_{0}+\frac{1}{2}(V_{\rm 1n}+V_{\rm 3n}) \\
E_{a}^{\theta =1/4}=V_{0}+3V_{\rm 3n} \\
E_{a}^{\theta =1/3}=V_{0}+3V_{\rm 2n} \\
E_{a}^{\theta =1/2}=V_{0}+V_{\rm 1n}+V_{\rm 2n}+3V_{\rm 3n}+V_{{\rm lt}} \\
E_{a}^{\theta =2/3}=V_{0}+\frac{3}{2}V_{\rm 1n}+3V_{\rm 2n}+\frac{3}{2}V_{\rm 3n}+
3V_{{\rm bt}} \\
E_{a}^{\theta =3/4}=V_{0}+2V_{\rm 1n}+2V_{\rm 2n}+3V_{\rm 3n}+2V_{{\rm lt}}+
2V_{{\rm bt}}+\frac{{}_{2}}{{}^{3}}V_{{\rm tt}} \\
E_{a}^{\theta =1.0}=V_{0}+3(V_{\rm 1n}+V_{\rm 2n}+V_{\rm 3n})+3V_{{\rm lt}}+
6V_{{\rm bt}}+2V_{{\rm tt}}
\end{array}
\]
where V$_{{\rm lt}}$, V$_{{\rm bt}}$ and V$_{{\rm tt}}$ are linear, bent and
triangular trios, respectively (indicated in Fig.~1).
We point out that only in the $(3 \times 3)$-2O structure at
$\theta =2/9$  can the atoms at nearest neighbor sites
move significantly from the 
locally 3-fold symmetric
adsorption sites to
reduce the repulsion, $V_{\rm 1n}$.
However, with strong nearest
neighbor repulsion, isolated nearest neighbor pairs are highly
improbable at any coverage. 
%%%%%%%%%%%%%%%%%%%%%%%%%%%%%%%%%%
%
% figure 1 should appear here
%
%%%%%%%%%%%%%%%%%%%%%%%%%%%%%%%%%%
\begin{figure}
\vspace{-10mm}
\psfig{figure=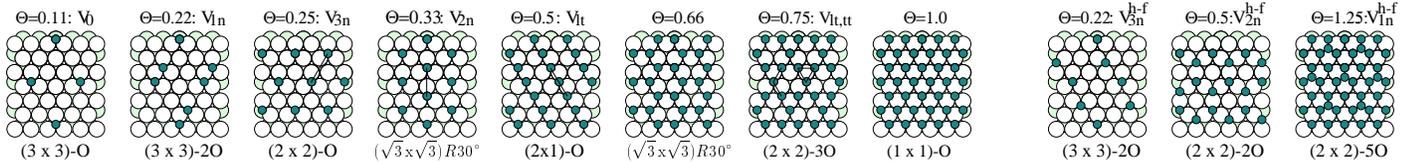,rheight=35mm}
\caption{Adsorbate structures calculated using DFT-GGA.
(For the first eight diagrams analogous calculations were also performed
for O in fcc sites.)
Small circles represent O atoms and large circles Ru atoms.}
\end{figure}                                
We have therefore calculated
$E_{a}^{\theta =2/9}$  for atoms at
locally 3-fold symmetric
sites because
if we were to use the relaxed structure we would need
to include higher many-body interactions (e.g. longer-ranged trios,
quartos and quintos
etc.) to account for the movement of the atoms back to the ideal
3-fold sites which occurs for higher coverages.

We thus have
for O atoms in hcp sites (and analogously for O atoms in fcc sites)
seven equations for six unknowns. Using the first six of
these equations we obtain the interaction energies listed in Tab.~II.
The accuracy of these interaction values is gauged by
calculating the adsorption energy of the monolayer (seventh equation); we 
find a discrepancy of
0.034~eV smaller than that of the DFT-GGA value. Along with our derived
values in Tab.~II, we give in brackets the interaction energies determined
by Piercy {\em et al.}~\cite{pdp} for their best-fit to the experimental
O/Ru\thinspace (0001) phase diagram. We also include our derived interaction
parameters for structures involving O atoms in 
hcp and fcc sites (last three diagrams of Fig.~1) obtained by
writing down appropriate equations in an analogous manner to those listed
above.

To complete the specification of our hamiltonian, we need the vibrational
frequencies of an O atom relative to the Ru surface. These frequencies can
also be calculated using density-functional theory. For example,
we obtain for the vibration of oxygen normal to the surface
(at the $\overline{\Gamma}$-point)
$\nu _{z} = 509$~cm$^{-1}$.  The experimental value is
535~cm$^{-1}$ (for O in the $(2 \times 2)$ phase)~\cite{o-vib}.

We now proceed to calculate
the temperature programmed thermal desorption spectra. Writing the kinetic
equation for adsorption and desorption as $d\theta /dt$ = $R_{{\rm ad}}-$
$R_{{\rm des}}$, we obtain for an atomic adsorbate in contact with a gas of
diatomic homonuclear molecules, the rate of adsorption $R_{{\rm ad}}$ =
$2S_{{\rm dis}}(\theta,T) P_{m}a_{s}\lambda _{m}/h$. Here $P_{m}$ is the
molecular pressure above the surface, $a_{s}$ is the area of one 
unit cell of the substrate surface,
$\lambda _{m}$ = $h/(2\pi mk_{B}T)^{1/2}$ is the thermal wavelength of
a molecule of mass $m$, and $S_{{\rm dis}}(\theta,T)$ is the dissociative
sticking coefficient. For the rate of desorption we have~\cite{kinrev},
\begin{equation}
R_{{\rm des}}=2S_{{\rm dis}}\left( \theta ,T\right)
a_{s}\frac{k_{B}T}{h\lambda _{m}^{2}}\frac{Z_{vr}}{q_{3}^{2}}{\frac{\theta
{}^{2}}{\left(
1-\theta \right) {}^{2}}}e^{\frac{-2|V_{0}|}{k_{B}T}}
e^{\frac{2\mu^{({\rm lat})}}{k_{B}T}}\quad .
\nonumber
\end{equation}
Here $Z_{vr}$ is the partition function accounting for the internal
vibrations and the rotations of O$_{2}$ in the gas phase.
$\mu ^{^{({\rm lat})}}$ is the contribution to the chemical

\newpage
potential of the adsorbate due to
the lateral interactions in the hamiltonian (Eq.~1) and is calculated here
using transfer matrix techniques.
Regarding the sticking coefficient,
we note that dissociation is not activated initially, but at (local)
coverages of
$\theta
\raisebox{.6ex}[-.6ex]{$>\!\!\!\!$}\raisebox{-.6ex}[.6ex]{$\sim$}\,
0.5$,
it is hindered by an energy
barrier~\cite{stampfl2,surnev}.
Under these circumstances, to obtain the 
coverage and temperature dependence of the
sticking coefficient 
{\em ab initio} would be a significant undertaking. 
Therefore we
use an analytic
expression which well approximates
the measured behavior in the
temperature regime of desorption~\cite{surnev}: The sticking coefficient
drops approximately as $\left( 2/3-\theta \right) {}^{2}$ and for coverages
above $2/3$ it remains very small up to a monolayer. The actual equation we
use is: $S_{{\rm dis}}(\theta )=S_{0}exp[-(\theta /\sigma )^{2}]$, with
$S_{0}$=0.27 and $\sigma $=0.3.
%%%%%%%%%%%%%%%%%%%%%%%%%%%%%%%%%%
%
% figure 2 should appear here
%
%%%%%%%%%%%%%%%%%%%%%%%%%%%%%%%%%%
\begin{figure}
\vspace{-10mm}
\psfig{figure=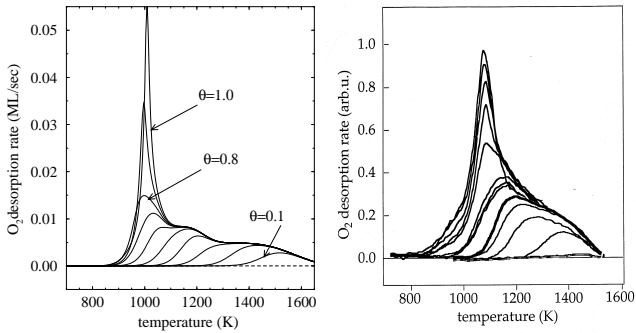,rheight=60mm}
\caption{Theoretical (left panel) and experimental~\protect\cite{boettcher}
(right panel) TPD spectra (heating rate of 6~K/s). For the
theoretical results initial coverages are $\Theta=$0.1 to 1.0 in
steps of 0.1; the experimental results also span the initial coverage region
of $\theta \rightarrow 0$ to 1ML.}
\end{figure}                                 

Our calculated TPD spectra~\cite{astek} are shown in Fig.~2 which
are compared
to recent experimental data~\cite{boettcher}. For low initial coverage
we note that in the theory the oxygen desorbs at about 100~K higher
temperature than in experiment, reflecting an overbinding of the O atoms.
We believe that this size of error is typical for present day
state-of-the-art calculations and don't see that (or how) better accuracy
will be achieved in the coming five to ten years.
Aside from this,
it can be seen that the theoretical spectra exhibit all the
features of the experimental data: (i) a shift of the peak maxima to lower
temperatures for higher initial coverages (appropriate for second order
desorption and/or for repulsive interactions) and (ii) a steepening of the
leading edge for higher initial coverages. In our theory this steepening is
a reflection of two facts: Firstly, for increasing coverages, the repulsive
\begin{table}
\begin{tabular}{c|cccccccc|c|ccc}
Site & $V_{0}=E_{a}^{\theta=1/9}$ & $E_{a}^{\theta=2/9}$ & $%
E_{a}^{\theta=1/4}$ & $E_{a}^{\theta=1/3}$ &
$E_{a}^{\theta=1/2}$ &$E_{a}^{\theta=2/3}$ & $%
E_{a}^{\theta=3/4}$ & $E_{a}^{\theta=1}$ &
Site & $E_{a,{\rm hcp-fcc}}^{\theta=2/9}$ & $E_{a,{\rm
hcp-fcc}}^{\theta=1/2}$ &
$E_{a,{\rm hcp-fcc}}^{\theta=5/4}$\\ \hline
hcp & $-$2.503 & $-$2.417 & $-$2.577 & $-$2.370 & $-$2.307 & $-$2.150 &
$-$2.091 & $-$%
1.895 & hcp-fcc & $-$2.294 & $-$2.209 & $-$1.492 \\
fcc & $-$2.152 & $-$2.107 & $-$2.145 & $-$2.105 &
$-$2.025 & $-$2.015 & $-$1.942 & $-$1.865 &   &   &  &   \\
\end{tabular}
\medskip
\caption{Adsorption energies (in eV) for
O on Ru\,(0001) with respect to 1/2O$_{2}$ for various coverages.}
\end{table}
\newpage
next nearest neighbor interactions lower the binding energy and thus lower
the onset of desorption, broadening the TPD spectra and steepening the
rising edge; 
secondly, and more importantly, is the effect of the rapidly
decreasing sticking probability for increasing coverage~\cite{o2ag}. 
Because
the sticking coefficient is much smaller for coverages above 2/3ML,
desorption is delayed to higher temperatures and the last third of a
monolayer desorbs over a very narrow temperature range. If we use the
ideal dissociative sticking, $S(\theta )=S_{0}(1-\theta )^{2}$, a
similar steepening still occurs but the onset of desorption occurs 100~K
earlier at a monolayer. A similar behavior has been discussed in
Ref.~\cite{o2ag}.
We can trace the two shoulders (at 1100K and 1300K) to the
synergy of the interactions that at lower temperatures lead to the formation
of the $(2 \times 1)$ and $(2 \times 2)$
ordered structures  which will also be seen in the
heat of adsorption.

Piercy {\em et al.} \cite{pdp} find that to obtain a satisfactory
explanation of the {\em surface phase diagram} in the vicinity of the
order-disorder transition temperature, a spillover into fcc sites of about
12~\% takes place. In the temperature regime of desorption, localization 
in
hcp sites should be even less. We find, however, that the overall 
features of the
TPD spectra remain essentially unchanged whether spillover is included or not.
We have also tested the importance of the trio interactions on desorption.
Neglecting them increases the overall repulsion for coverages larger than
2/3~ML and consequently broadens the TPD spectra, reducing the
agreement with experiment. We therefore conclude that for high O-coverages
trio interactions play an important role.
%%%%%%%%%%%%%%%%%%%%%%%%%%%%%%%%%%
%
% figure 3 should appear here
%
%%%%%%%%%%%%%%%%%%%%%%%%%%%%%%%%%%
\begin{figure}
\vspace{-10mm}
\psfig{figure=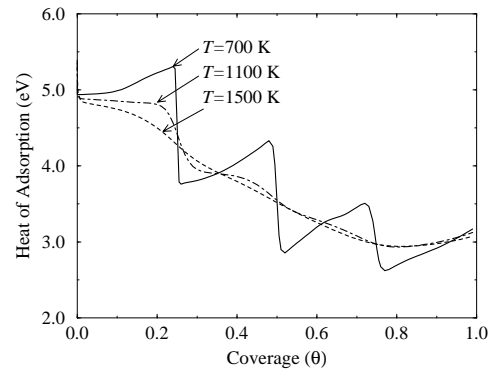,rheight=65mm}
\caption{The heat of adsorption as a function of coverage for temperatures
of $T$=700~K (sharpest features), 1100~K, and 1500~K.}
\end{figure}

\newpage
Having the chemical potential as a function of temperature and coverage, we
can also calculate the equilibrium properties of the adsorbate, such as
adsorption isobars. As an example, we present in Fig.~3 the isosteric heat
of adsorption~\cite{kinrev} for a few temperatures.
At the highest temperature a smooth
decrease is observed. At the lowest temperature, sharp peaks (and decreases) at 1/4,
1/2, 3/4, and 1~ML occur. These coverages correspond to each
of the ordered
phases that form {\em in nature}, i.e.
$(2 \times 2)$-O~\cite{lindroos}, $(2 \times 1)$-O~\cite{P1}, $(2 \times
2)$-3O~\cite
{kostov,over,menzel}, and $(1 \times 1)$-O~\cite{stampfl2}. 
(Note that the existence of the $(2 \times 2)$-3O, and $(1 \times 1)$-O
adsorption phases was at first predicted by DFT-GGA calculations \cite{stampfl1}
and subsequently confirmed by experiment; a nice success of DFT.)
The rises in
between originate from the third neighbor attractions, and also from the
trios for the higher coverages. Our results show no tendency for the
stability of a $\theta=1/3$ phase, i.e. a $(\sqrt{3} \times
\sqrt{3})R30^{\circ}$ structure, in agreement with experiment.
We now turn to the surface phase diagram. From Tab.~II it can be seen that
the overall agreement of the interaction parameters determined from our
density-functional calculations and from the best-fit to experiment in
Ref.~\cite{pdp} is, in general, astonishingly good; but there are some  
significant differences (i.e. more than 50\% for $V_{\rm 1n}^{\rm fcc}$)
which could be expected. 
We find that
our Monte Carlo simulations yields a
surface phase diagram rather similar to that of Piercy {\em et al.}~\cite
{pdp} where coverage up to half a monolayer was considered. Our present
simulations
also included higher O coverages
and correctly predict formation of the $(2 \times 2)$-3O phase.

In summary, we have presented a first-principles based approach for
calculation of the thermodynamics and kinetics of an adsorbate on a surface.
We used density-functional theory to create a lattice gas hamiltonian from
which we evaluated the partition function. Our theoretical temperature
programmed thermal desorption spectra, heats of adsorption, and the surface
phase diagram for O on Ru\,(0001) show very good overall agreement with
available experimental results, providing confidence in our approach. We
found that trio interactions, as well as the sticking coefficient, play an
important role in the TPD spectra of the present system. The attractive trio
interactions also apparently help stabilize the higher coverage $(2 \times
1) $-O, $(2 \times 2)$-3O, and $(1 \times 1)$-O phases. The affect of
spillover into fcc sites was found to have a minimal affect on the TPD
spectra.
\begin{table}
\begin{tabular}{c|cccccc}
Site & $V_{\rm 1n}$ & $V_{\rm 2n}$ &
$V_{\rm 3n}$ & $V_{{\rm lt}}$ & $V_{{\rm bt}}$ &
$V_{{\rm tt}}$ \\ \hline
hcp & 0.265 & 0.044 & $-$0.025 & $-$0.039 & $-$0.046 & 0.058 \\
& (0.23) & (0.069) & ($-$0.023) &  &  &  \\ \hline
fcc & 0.158 & 0.016 & 0.002 & $-$0.052 & $-$0.044 & 0.076 \\
& (0.069) &  &  &  &  &  \\ \hline
fcc-hcp & 0.586 & 0.101 & 0.033 &  &  &
\end{tabular}
\medskip
\caption{DFT-GGA calculated interaction energies (in eV)
Piercy {\em et al.}~\protect\cite{pdp}.}
\end{table}

\end{document}